\documentclass[11pt,twoside]{article}
\input{epsf.sty}
\usepackage{mathrsfs}
\usepackage{amsmath}
\usepackage{amssymb}
\usepackage{graphicx}
\usepackage{subfigure}
\newtheorem{proposition}{Proposition}

\DeclareMathOperator{\sech}{sech}

\title{Darboux transformation and multi-soliton solutions of the Camassa-Holm equation and modified Camassa-Holm equation}
\author{Baoqiang Xia$^{1}$\footnote{E-mail address:
xiabaoqiang@126.com}, Ruguang Zhou$^1$\footnote{E-mail address:
zhouruguang@jsnu.edu.cn}, Zhijun Qiao$^2$\footnote{ E-mail address:
qiao@utpa.edu},
\\ $^{1}$School of Mathematics and Statistics, Jiangsu Normal
University,\\
 Xuzhou, Jiangsu 221116, P. R. China
 \\
$^2$Department of Mathematics, University of Texas-Pan American, \\Edinburg, Texas 78541, USA}

\date{}
\begin{document}
\maketitle
\begin{abstract}

In this paper, we propose a new approach to calculate multi-soliton solutions of Camassa-Holm (CH) equation and modified Camassa-Holm (MCH) equation
with aid of Darboux transformation (DT). The new approach simplifies the approach presented in {\it Proc. R. Soc. Lond. A} {\bf 460} 2617-2627 (2004).
We first map the CH and MCH equation to a negative order KdV (NKdV) equation by a reciprocal transformation.
Then we proceed to apply the DT to solve the NKdV equation in the usual way.
Finally we invert the reciprocal transformation to recover the solutions of the CH equation and MCH equation.

\vspace*{0.2cm}

\end{abstract}
\newpage

\section{Introduction}

In recent years, the Camassa-Holm (CH) equation \cite{CH}
\begin{eqnarray}
q_t+2q u_x+q_xu=0, \quad q=m+k^2=u-u_{xx}+k^2,
\label{CH}
\end{eqnarray}
($k$ is an arbitrary constant) derived by Camassa and Holm \cite{CH} as a shallow water
wave model, has attracted much attention and various studies.
This equation first appeared in the paper of Fuchssteiner and Fokas \cite{FF1} on hereditary symmetries, but it came to be remarkable in the
work of Camassa and Holm \cite{CH} where the peakon was described.
As an integrable system, the CH equation shares most of the important properties of an integrable
system of KdV type. For example, it admits Lax representation \cite{CH}, bi-Hamiltonian formula \cite{CH,OR}, and can be solved by the inverse scattering transformation \cite{BSZ,C3}.
The most remarkable feature of the CH equation is that it admits peaked
soliton (peakon) solutions in the case $k=0$ \cite{CH}.
A peakon is a weak solution in some Sobolev space with corner at its crest.
The stability of the peakon were proved by Constantin and Strauss in the references \cite{CS1,CS2}.

The interesting  features of the CH equation have excited many endeavors to seek smooth soliton solutions of the CH equation.
By a loop group approach, Schiff \cite{JS} found a single-soliton solution in a parametric form and an explicit
(but incomplete) expression for the two-soliton solution.
In the reference \cite{C3}, Constantin presented a complete description of the scattering and inverse scattering problems associated with the CH equation.
The key observation made by Constantin is that the isospectral problem of CH equation was mapped onto a Schr\"{o}dinger spectral problem by introducing the Liouville transformation.
Latter, Johnson \cite{JR} implemented the inverse scattering method for CH equation developed by Constantin to derive the soliton solutions of CH equation.
Johnson succeeded in finding single-, two- and three-soliton solutions, but encountered difficulties in seeking the multi-soliton
of the CH equation, since his procedure relies on a good guesswork and an extensive use of mathematical packages.
To overcome the difficulties, Parker \cite{AP1,AP2,AP3} presented a different route to derive multi-soliton solutions of the CH equation based on Hirota's bilinear transformation method.
Parker first showed that the CH equation can be mapped onto a version of the AKNS-SWW equation \cite{AKNS} by using the reciprocal transformation proposed in \cite{JS,Fu}.
Then Parker solved the AKNS-SWW equation by Hirota's bilinear transformation method. Finally by the inverse reciprocal transformation, Parker arrived at the desired multi-soliton solutions of the CH equation.

In this paper, inspired by the Parker's work \cite{AP1,AP2,AP3}, we propose an approach to
compute the multi-soliton solution of CH equation with aid of Darboux transformation (DT). We remark that in \cite{LZ}, Li and Zhang have refined Johnson's method \cite{JR}
for calculating explicit soliton solutions of the Camassa-Holm equation via DT.
They found an analytical formula for the potential $q$ and presented an algorithm for solving for the potential $u$.
The Li-Zhang's approach is interesting and instructive to us, but the approach did not give an explicit formula for the multi-soliton solution for the potential $u$,
since the algorithm requires one to solve second-order ordinary differential equations (ODEs) to compute the solution $u$.
In contrast with Li-Zhang's approach, our approach provides a more direct and simple way (avoid having to solve the ODEs encountered by Li and Zhang)
to calculate the multi-soliton solution for the potential $u$ of CH equation.
We first make a connection between the CH equation and a negative order KdV (NKdV) equation by the reciprocal transformation proposed in \cite{JS,Fu}.
Then we proceed to apply the DT to solve the NKdV equation in the usual way.
Based on the results of the DT, we present a procedure to invert the reciprocal transformation to recover the solution of the potential $u$.

In this paper, we also show our approach can be applied to derive the multi-soliton solution of the modified CH (MCH) equation with cubic nonlinearity \cite{OR,Fo,Fu,Q1,GLOQ}
\begin{eqnarray}
 m_t+\left[ m(u^2-u^2_x)\right]_x=0, \quad  m=u-u_{xx}.\label{mCH}
\end{eqnarray}
Similar with the case of the CH equation, we first map the MCH equation onto a NKdV equation by a reciprocal transformation.
Then by using our procedure, we arrive at the multi-soliton solution of MCH equation.

The whole paper is organized as follows. In Section 2, we shortly summarize the reciprocal transformation of CH equation,
and describe our approach for computing the multi-soliton solution of CH equation.
Then in Section 3 and Section 4, we apply our approach to calculate the multi-soliton solutions of CH equation and MCH equation, respectively.
Some conclusions and discussions are addressed in Section 5.

\section{A reciprocal transformation for CH equation}

In this section, we first review some basic results about the reciprocal transformation of CH equation presented in \cite{C3,JS,Fu}.
These results enable us to relate CH equation to a NKdV equation.
Then we propose our approach to calculate the multi-soliton solution of CH equation with aid of DT at the end of this section.

It was shown in \cite{CH} that the CH equation possesses the following Lax pair with the spatial part
\begin{eqnarray}
\psi_{xx}=(\frac{1}{4}+\lambda q)\psi,
\label{lpsch}
\end{eqnarray}
and the temporal part
\begin{eqnarray}
\psi_t=(\frac{1}{2\lambda}-u)\psi_x+\frac{1}{2}u_x\psi,
\label{lptch}
\end{eqnarray}
where
$\lambda$ is a spectral parameter.

The Liouville transformation \cite{C3}
\begin{eqnarray}
r=\sqrt{q}=\sqrt{m+k^2},
\label{lt}
\end{eqnarray}
converts the CH equation into conservation form
\begin{eqnarray}
r_t+(ur)_x=0.
\label{ch2}
\end{eqnarray}
This permits us to define a reciprocal transformation $(x,t)$ $\mapsto$  $(y,\tau)$ by the relation
\begin{eqnarray}
dy=rdx-urdt, \quad d\tau=dt,
\label{rt1}
\end{eqnarray}
and we have
\begin{eqnarray}
\frac{\partial}{\partial x}=r\frac{\partial}{\partial y}, \quad \frac{\partial}{\partial t}=\frac{\partial}{\partial \tau}-ur\frac{\partial}{\partial y}.
\label{ct1}
\end{eqnarray}
This reciprocal transformation was originally suggested in \cite{Fu} and was later taken up in \cite{JS,C3,JR} and \cite{LZ,AP1}.

Under the reciprocal transformation (\ref{rt1}), the CH equation (\ref{CH}) is transformed to a coupled system
\begin{eqnarray}
&&r_{\tau}+r^2u_y=0,
\label{aCH}
\\
&&u=r^2-r\partial^2_{y\tau}\ln r-k^2, \label{u1}
\end{eqnarray}
which was called as associated CH (ACH) equation \cite{JS,H1}.

Denote $\phi=r^{\frac{1}{2}}\psi$. By using the reciprocal transformation (\ref{rt1}), the spatial part of the  spectral problem (\ref{lpsch}) becomes a Schr\"{o}dinger spectral problem
\begin{eqnarray}
\phi_{yy}=\lambda\phi+U\phi, \label{lpsch2}
\end{eqnarray}
with the potential
\begin{eqnarray}
U=\frac{1}{2}\frac{r_{yy}}{r}-\frac{1}{4}\frac{r_y^2}{r^2}+\frac{1}{4r^2}. \label{U1}
\end{eqnarray}
Now Li-Zhang's procedure \cite{LZ} for solving the CH equation using the DT is apparent.
They first construct the potential $U$ by applying the DT to the Schr\"{o}dinger spectral problem (\ref{lpsch2}).
Then they solve (\ref{U1}) for $r$ with the known $U$ and solve $u-u_{xx}+k^2=r^2$ for $u$ with the known $r$.
By integrating $\frac{dy}{dx}=r$, they build up the relation between $x$ and $y$ with the known $r$.
Finally, they introduce a variable $t$ in the solution $u$ of the CH equation.
The Li-Zhang's procedure is interesting and very instructive to us,
but the difficulty encountered in executing the procedure is that one have to solve the ODE (\ref{U1}) and the ODE $u-u_{xx}+k^2=r^2$ to complete the solution $u$.
Inspired by the Parker's work \cite{AP1}, in the following we will accomplish the solution $u$ of CH equation using DT by a quite different procedure, which enables us to avoid solving the ODEs encountered in Li-Zhang's procedure.

We notice that the reciprocal transformation (\ref{rt1}) casts the temporal part of the spectral problem (\ref{lptch}) into
\begin{eqnarray}
\phi_{\tau}=\frac{1}{2\lambda}(r\phi_y-\frac{1}{2}r_y\phi). \label{lptch2}
\end{eqnarray}
The compatibility condition of (\ref{lpsch2}) and (\ref{lptch2}) gives rise to
\begin{eqnarray}
\left\{\begin{array}{l}
U_{\tau}=r_y,
\\
r_{yyy}-4Ur_y-2rU_y=0.
\end{array}\right.
\label{nkdv1}
\end{eqnarray}
We point out that (\ref{nkdv1}) is nothing but a negative order KdV (NKdV) equation \cite{Z,H1,Zhou1,Zhou2,QF}.

Armed with the above results, we are now able to present a simple approach to derive the multi-soliton solution of CH equation via DT. The procedure is constituted of the following three steps:
\begin{itemize}

\item apply Darboux transformation to NKdV equation (\ref{nkdv1}) to derive the solutions $U$ and $r$;

\item substitute the known $r$ into the formula (\ref{u1}) to calculate the potential $u$;

\item build up a relation between the variables $x$, $t$ and $y$, $\tau$.
\end{itemize}

\section{ Application to solving CH equation}

Now we execute the procedure we presented above to calculate the multi-soliton solutions of CH equation.
\subsection*{Step 1: Darboux transformation of the NKdV equation (\ref{nkdv1})}

Let $\phi_0 = \phi_0(y,\tau)$ be a solution of Lax pair (\ref{lpsch2}) and (\ref{lptch2})
fixed at the point $\lambda=\lambda_0$, and use it to define the gauge transformation
\begin{equation}
\tilde{\phi}=(\partial_y-\sigma)\phi,
\label{gt}
\end{equation}
where $\sigma=\partial_y\ln\phi_0$.
Let us set
\begin{equation}
\tilde{U}=U-\sigma_{y}, \quad \tilde{r}=r-\sigma_{\tau}.
\end{equation}
By a direct calculation, we may check that (for the details see \cite{QF}) the Lax pair (\ref{lpsch2}) and (\ref{lptch2}) is covariant with respect to the action of the Darboux
transformation
\begin{equation}
\phi\rightarrow \tilde{\phi}, \quad U\rightarrow \tilde{U}, \quad r\rightarrow \tilde{r}.
\label{dt}
\end{equation}
Thus we arrive at the following proposition.
\begin{proposition}
A solution $U$, $r$ of the NKdV equation (\ref{nkdv1})
is mapped onto its new solution $\tilde{U}$, $\tilde{r}$ under the Darboux
transformation (\ref{dt}).
\end{proposition}

Now we apply the above Darboux transformation to derive the multi-soliton solutions for the NKdV equation (\ref{nkdv1}).
Let $f_1,f_2,\cdots,f_N$ be solutions of (\ref{lpsch2}) and (\ref{lptch2}) fixed at $\lambda=\lambda_1,\lambda_2,\cdots,\lambda_N$.
We start with a seed solution $r=k$ and $U=\frac{1}{4k^2}$ and introduce the parameters $p_j$, $c_j$ by $\frac{1}{4}p_j^2=\lambda_j+\frac{1}{4k^2}$, $c_j=\frac{2k^3}{k^2p_j^2-1}$.
Then the functions $f_1,f_2,\cdots,f_N$ can be chosen as
\begin{eqnarray}
f_j=\left\{\begin{array}{l}
cosh \xi_j, \quad j=2l-1,
\\
sinh \xi_j, \quad j=2l,
\end{array}\right.
\label{fj}
\end{eqnarray}
where $\xi_j=\frac{1}{2}p_j(y+c_j\tau)$.
We define the Wronskian determinant $W_N$ of $N$ functions $f_1,f_2,\cdots,f_N$ by
\begin{eqnarray}
W_N=det A, \quad A_{ij}=\frac{d^{i-1}f_j}{dy^{i-1}}, \quad i,j=1,2,\cdots,N.\label{WT}
\end{eqnarray}
By repeating the Darboux transformation $N$-times (see the reference \cite{MS} for example),
the $N$-soliton solutions of the NKdV equation (\ref{nkdv1}) can be expressed as
\begin{eqnarray}
U&=&\frac{1}{4k^2}-2\frac{\partial^2}{\partial y^2}\ln W(f_1,f_2,\cdots,f_N)\triangleq \frac{1}{4k^2}-2\frac{\partial^2}{\partial y^2}\ln W_N,
\label{usch}
\\
r&=&k-2\frac{\partial^2}{\partial y \partial \tau}\ln W(f_1,f_2,\cdots,f_N)\triangleq k-2\frac{\partial^2}{\partial y \partial \tau}\ln W_N.
\label{rsch}
\end{eqnarray}
The two fundamental solutions of equation (\ref{lpsch2}) with $\lambda=0$ and $U$ given by (\ref{usch}) are
\begin{eqnarray}
\left\{\begin{array}{l}
\phi_1=\frac{W(f_1,f_2,\cdots,f_N,e^{\frac{1}{2k}y})}{W(f_1,f_2,\cdots,f_N)}\triangleq \frac{\bar{W}_N}{W_N},
\\
\phi_2=\frac{W(f_1,f_2,\cdots,f_N,e^{-\frac{1}{2k}y})}{W(f_1,f_2,\cdots,f_N)}\triangleq \frac{\hat{W}_N}{W_N}.
\end{array}\right.
\label{phich}
\end{eqnarray}
Their asymptotic properties are
\begin{eqnarray}
\begin{array}{l}
\phi_1\sim \prod\limits_{j-1}^{N}(\frac{1}{2k}-\frac{p_j}{2})e^{\frac{1}{2k}y},
\quad
\phi_2\sim \prod\limits_{j-1}^{N}(-\frac{1}{2k}-\frac{p_j}{2})e^{-\frac{1}{2k}y},
\\
\phi_{1,y}\sim \frac{1}{2k}\prod\limits_{j-1}^{N}(\frac{1}{2k}-\frac{p_j}{2})e^{\frac{1}{2k}y},
\quad
\phi_{2,y}\sim -\frac{1}{2k}\prod\limits_{j-1}^{N}(-\frac{1}{2k}-\frac{p_j}{2})e^{-\frac{1}{2k}y}.
\end{array}
\label{ap1}
\end{eqnarray}

\subsection*{Step 2: the solution $u(y,\tau)$ of CH equation}

The formula (\ref{rsch}) provides a multi-soliton solution of the ACH equation (\ref{aCH}).
Substituting (\ref{rsch}) into (\ref{u1}) yields
\begin{eqnarray}
u=4[(\ln W_N)_{y\tau}]^2-4k(\ln W_N)_{y\tau}-[k-2(\ln W_N)_{y\tau}][\ln (k-2(\ln W_N)_{y\tau})]_{y\tau}, \label{us1}
\end{eqnarray}
which provides an analytic solution of the CH equation expressed in the variables $y$ and $\tau$.

\subsection*{Step 3: the relation between $x$, $t$ and $y$, $\tau$}
To complete the solution $u(x,t)$ of the CH equation, we have only to obtain the
coordinate transformation between $x$, $t$ and $y$, $\tau$. There are a number of ways to
accomplish this final step (for example, see \cite{JS,JR,LZ,AP1}). We finish this step based on a result of the DT presented in Step 1.
When $\lambda=0$, equations (\ref{lpsch}) and (\ref{lpsch2}) become
\begin{eqnarray}
\psi_{xx}=\frac{1}{4}\psi, \label{ps2}
\\
\phi_{yy}=U\phi. \label{ph2}
\end{eqnarray}
For equation (\ref{ps2}), we may choose a solution $\psi$ to be $\psi=e^{\frac{x}{2}}$.
For equation (\ref{ph2}), the solution $\phi$ is a linear combination of the two fundamental solutions $\phi_1$ and $\phi_2$ given by (\ref{phich}).
The asymptotic behaviours (\ref{ap1}) imply that the $\phi$ corresponding to $\psi=e^{\frac{x}{2}}$ must be $d_1\phi_1$, where $d_1$ is a constant.
Thus the relation $\phi=r^{\frac{1}{2}}\psi$ yields
\begin{eqnarray}
e^{\frac{x}{2}}=\frac{d_1\phi_1}{\sqrt{r}}, \label{xy}
\end{eqnarray}
which gives rise to the following relation
\begin{eqnarray}
x=\ln\frac{d_1^2\phi_1^2}{r}=\ln\frac{ \bar{W}^2_N}{[k-2(\ln W_N)_{y\tau}]W^2_N}+\alpha, \label{xyr}
\end{eqnarray}
where $\alpha$ is a constant.
On the other hand, from (\ref{rt1}) it is easy to see
\begin{eqnarray}
t=\tau. \label{xyr1}
\end{eqnarray}
The formulas (\ref{xyr}) and (\ref{xyr1}) constitute the desired relation between $x$, $t$ and $y$, $\tau$.

\subsection*{Examples: one-soliton solution and two-soliton solution}

The formula (\ref{us1}) together with (\ref{xyr}) and (\ref{xyr1}) provides an analytic solitary solution of the CH equation in parametric form.
As examples, we present the single-soliton solution and two-soliton solution in the following.

\subsubsection*{Example 1: one-soliton solution of CH equation}
For $N=1$, from (\ref{fj}) we take
\begin{eqnarray}
f_1=\cosh \xi_1, \quad \xi_1=\frac{1}{2}p_1(y+c_1\tau),
\label{f1ch}
\end{eqnarray}
where $\frac{1}{4}p_1^2=\lambda_1+\frac{1}{4k^2}$, $c_1=\frac{2k^3}{k^2p_1^2-1}$.
The potential function $U$ given by the general formula (\ref{usch}) becomes
\begin{eqnarray}
U=\frac{1}{4k^2}-2(\ln f_1)_{yy}=\frac{1}{4k^2}-\frac{1}{2}p_1^2\sech^2\xi_1. \label{Uscho}
\end{eqnarray}
From the general formula (\ref{rsch}), we obtain the one-soliton solution of ACH equation
\begin{eqnarray}
r=k-2(\ln f_1)_{y\tau}=k-\frac{1}{2}c_1p_1^2\sech^2\xi_1.
\label{rs1cho}
\end{eqnarray}
From the general formulas (\ref{us1}), (\ref{xyr}) and (\ref{xyr1}),
we arrive at the one-soliton solution of CH equation in parametric form
\begin{eqnarray}
\begin{split}
u&=4[(\ln f_1)_{y\tau}]^2-4k(\ln f_1)_{y\tau}-[k-2(\ln f_1)_{y\tau}][\ln (k-2(\ln f_1)_{y\tau})]_{y\tau}
\\&=\frac{kc_1p_1^2(c_1p_1^2-2k)\sech^2\xi_1}{-c_1p_1^2\sech^2\xi_1+2k},
\\
x&=\frac{y}{k}+\ln\frac{(\frac{1}{k}\cosh\xi_1-p_1\sinh\xi_1)^2}{(k-\frac{1}{2}c_1p_1^2\sech^2\xi_1)\cosh^2\xi_1}+\alpha
\\&=\frac{y}{k}+\ln\frac{(1-k^2p_1^2)[(1+kp_1)+(1-kp_1)e^{2\xi_1}]}{k^3[(1-kp_1)+(1+kp_1)e^{2\xi_1}]}+\alpha,
\\
t&=\tau.
\end{split}
\label{ucho}
\end{eqnarray}
This parametric representation of the one-soliton solution of CH equation is first found in \cite{JS} and latter confirmed in \cite{JR,LZ,AP1}.

\subsubsection*{Example 2: two-soliton solution of CH equation}

For $N=2$, we take
\begin{eqnarray}
f_1=\cosh \xi_1, \quad f_2=\sinh\xi_2, \quad \xi_1=\frac{1}{2}p_1(y+c_1\tau), \quad \xi_2=\frac{1}{2}p_2(y+c_2\tau),
\label{f2ch}
\end{eqnarray}
where $\frac{1}{4}p_1^2=\lambda_1+\frac{1}{4k^2}$, $c_1=\frac{2k^3}{k^2p_1^2-1}$, $\frac{1}{4}p_2^2=\lambda_2+\frac{1}{4k^2}$, $c_2=\frac{2k^3}{k^2p_2^2-1}$.
The potential function $U$ given by the general formula (\ref{usch}) is cast into
\begin{eqnarray}
U=\frac{1}{4k^2}-2(\ln W_2)_{yy}
 =\frac{1}{4k^2}-\frac{(p_2^2-p_1^2)(p_2^2\cosh^2\xi_1+p_1^2\sinh^2\xi_2)}{2(p_2\cosh\xi_1\cosh\xi_2-p_1\sinh\xi_1\sinh\xi_2)^2}.
 \label{Uscht}
\end{eqnarray}
Based on the general formula (\ref{rsch}), the two-soliton solution of ACH equation reads
\begin{eqnarray}
r=k-2(\ln W_2)_{y\tau}
 =k-\frac{(p_2^2-p_1^2)(p_2^2c_2\cosh^2\xi_1+p_1^2c_1\sinh^2\xi_2)}{2(p_2\cosh\xi_1\cosh\xi_2-p_1\sinh\xi_1\sinh\xi_2)^2}
 \triangleq k-\Gamma.
\label{rs1cht}
\end{eqnarray}
From the general formulas (\ref{us1}), (\ref{xyr}) and (\ref{xyr1}),
we arrive at the two-soliton solution of CH equation in parametric form
\begin{eqnarray}
\begin{split}
u&=4[(\ln W_2)_{y\tau}]^2-4k(\ln W_2)_{y\tau}-[k-2(\ln W_2)_{y\tau}][\ln (k-2(\ln W_2)_{y\tau})]_{y\tau}
\\&=\Gamma^2-2k\Gamma+\Gamma_{y\tau}+\frac{\Gamma_y\Gamma_{\tau}}{k-\Gamma},
\\
x&=\ln\frac{\bar{W}^2_2}{[k-2(\ln W_2)_{y\tau}]W^2_2}+\alpha
\\&=\frac{y}{k}+\ln\frac{\left[p_1(p_2^2-\frac{1}{k^2})\sinh\xi_1\sinh\xi_2-p_2(p_1^2-\frac{1}{k^2})\cosh\xi_1\cosh\xi_2
+\frac{1}{k}(p_1^2-p_2^2)\cosh\xi_1\sinh\xi_2\right]^2}{16k(p_2\cosh\xi_1\cosh\xi_2-p_1\sinh\xi_1\sinh\xi_2)^2
-8(p_2^2-p_1^2)(p_2^2c_2\cosh^2\xi_1+p_1^2c_1\sinh^2\xi_2)}+\alpha,
\\
t&=\tau,
\end{split}
\label{ucht}
\end{eqnarray}
where the function $\Gamma$ is given in the formula (\ref{rs1cht}).
By a reformulation, this solution can be made to agree with the two-soliton solutions reported in \cite{JS,JR,LZ,AP2}.

\section{ Application to solving MCH equation}

In this section, we shall apply our procedure to calculate the multi-soliton solution of the MCH equation (\ref{mCH}).
Similar with the CH equation case, we first map the MCH equation onto NKdV equation by a reciprocal transformation.

\subsection*{A Reciprocal transformation for MCH equation}

The MCH equation possesses the following Lax pair with the spatial part \cite{Q1}
\begin{eqnarray}
\left\{\begin{array}{l}
\psi_x=-\frac{1}{2}\psi+\frac{1}{2}\lambda m\phi,
\\
\phi_x=-\frac{1}{2}\lambda m\psi+\frac{1}{2}\phi,
\end{array}\right.
\label{lps}
\end{eqnarray}
and the temporal part
\begin{eqnarray}
\left\{\begin{array}{l}
\psi_t=[\lambda^{-2}+\frac{1}{2}(u^2-u^2_x)]\psi-[\lambda^{-1}(u-u_x)+\frac{1}{2}\lambda m(u^2-u^2_x)]\phi,
\\
\phi_t=[\lambda^{-1}(u+u_x)+\frac{1}{2}\lambda m(u^2-u^2_x)]\psi-[\lambda^{-2}+\frac{1}{2}(u^2-u^2_x)]\phi,
\end{array}\right.
\label{lpt}
\end{eqnarray}
where
$\lambda$ is a spectral parameter.

We define a reciprocal transformation $(x,t)$ $\mapsto$  $(y,\tau)$ by the relation \cite{IL,MY}
\begin{eqnarray}
dy=mdx-m(u^2-u^2_x)dt, \quad d\tau=dt.
\label{rt}
\end{eqnarray}
Then we have
\begin{eqnarray}
\frac{\partial}{\partial x}=m\frac{\partial}{\partial y}, \quad \frac{\partial}{\partial t}=\frac{\partial}{\partial \tau}-m(u^2-u^2_x)\frac{\partial}{\partial y}.
\label{ct}
\end{eqnarray}
Under the reciprocal transformation (\ref{rt}), the MCH equation (\ref{mCH}) is transformed to
\begin{eqnarray}
 m_{\tau}+2m^3u_y=0, \label{tmCH}
\end{eqnarray}
and the potential $u$ can be expressed in terms of $m$ as
\begin{eqnarray}
u=m+\frac{1}{2}m(\frac{1}{m})_{\tau y}. \label{u}
\end{eqnarray}
We call the coupled system (\ref{tmCH}) and (\ref{u}) as associated MCH (AMCH) equation \cite{MY}.

By the reciprocal transformation (\ref{rt}), the spatial part of the  spectral problem (\ref{lps}) is cast into a Schr\"{o}dinger spectral problem
\begin{eqnarray}
\phi_{yy}=-\frac{\lambda^2}{4}\phi+U\phi, \label{lps1}
\end{eqnarray}
with the potential
\begin{eqnarray}
U=\frac{1}{4m^2}-\frac{m_y}{2m^2}. \label{U}
\end{eqnarray}
Moreover, the temporal part of the spectral problem (\ref{lpt}) is cast into
\begin{eqnarray}
\phi_{\tau}=-\frac{1}{2\lambda^2}V_y\phi+\frac{1}{\lambda^2}V\phi_y, \label{lpt1}
\end{eqnarray}
with
\begin{eqnarray}
V=-2(u+u_x)=-2(u+mu_y). \label{V}
\end{eqnarray}
The compatibility condition of (\ref{lps1}) and (\ref{lpt1}) gives rise to
\begin{eqnarray}
\left\{\begin{array}{l}
U_{\tau}=-\frac{1}{2}V_y,
\\
V_{yyy}-2VU_y-4V_yU=0,
\end{array}\right.
\label{nkdv2}
\end{eqnarray}
which is nothing but a negative order KdV equation.

Let $\rho=\frac{\phi_y}{\phi}$. From (\ref{lps1}), we can easily verify that  $\rho$ satisfies the Riccati equation
\begin{eqnarray}
\rho_y=-\frac{\lambda^2}{4}+U-\rho^2.
\label{RE}
\end{eqnarray}
Inserting (\ref{U}) into (\ref{RE}), we can conclude that \cite{IL}
\begin{eqnarray}
m=\left.\frac{1}{2\rho}\right|_{\lambda=0}=\left.\frac{\phi(y,\tau,\lambda)}{2\phi_y(y,\tau,\lambda)}\right|_{\lambda=0}.
\label{mphi}
\end{eqnarray}
The formula (\ref{mphi}) will be used to calculate the multi-soliton solution of the AMCH equation.

Prepared with the above results, we now can execute our procedure to compute the soliton solutions of MCH equation.

\subsection*{Step 1: Darboux transformation of the NKdV equation (\ref{nkdv2})}
We start with a seed solution $V=-2k$ and $U=\frac{1}{4k^2}$ of NKdV equation (\ref{nkdv2}). Similar with the step 1 in the above section,  by applying the Darboux transformation, the $N$-soliton solution of NKdV equation (\ref{nkdv2}) can be expressed as
\begin{eqnarray}
\begin{split}
U&= \frac{1}{4k^2}-2\frac{\partial^2}{\partial y^2}\ln W_N,
\\
V&=-2k+4\frac{\partial^2}{\partial y\tau}\ln W_n,
\end{split}
\label{uvmch}
\end{eqnarray}
where $W_N$ is the Wronskian determinant of $N$ functions $f_1,f_2,\cdots,f_N$.
The functions $f_j$ are defined by the formulas
\begin{eqnarray}
f_j=\left\{\begin{array}{l}
cosh \xi_j, \quad j=2l-1,
\\
sinh \xi_j, \quad j=2l,
\end{array}\right.
\label{fj2}
\end{eqnarray}
where $\xi_j=\frac{1}{2}p_j(y+c_j\tau)$ with $\frac{1}{4}p_j^2=-\frac{1}{4}\lambda_j^2+\frac{1}{4k^2}$, $c_j=\frac{2k^3}{k^2p_j^2-1}$.
The two fundamental solutions of equation (\ref{lps1}) with $\lambda=0$ and $U$ chosen as in (\ref{uvmch}) are
\begin{eqnarray}
\left\{\begin{array}{l}
\phi_1=\frac{W(f_1,f_2,\cdots,f_N,e^{\frac{1}{2k}y})}{W(f_1,f_2,\cdots,f_N)}\triangleq \frac{\bar{W}_N}{W_N},
\\
\phi_2=\frac{W(f_1,f_2,\cdots,f_N,e^{-\frac{1}{2k}y})}{W(f_1,f_2,\cdots,f_N)}\triangleq \frac{\hat{W}_N}{W_N}.
\end{array}\right.
\label{mchp}
\end{eqnarray}
Their asymptotic properties are
\begin{eqnarray}
\begin{array}{l}
\phi_1\sim \prod\limits_{j-1}^{N}(\frac{1}{2k}-\frac{p_j}{2})e^{\frac{1}{2k}y},
\quad
\phi_2\sim \prod\limits_{j-1}^{N}(-\frac{1}{2k}-\frac{p_j}{2})e^{-\frac{1}{2k}y},
\\
\phi_{1,y}\sim \frac{1}{2k}\prod\limits_{j-1}^{N}(\frac{1}{2k}-\frac{p_j}{2})e^{\frac{1}{2k}y},
\quad
\phi_{2,y}\sim -\frac{1}{2k}\prod\limits_{j-1}^{N}(-\frac{1}{2k}-\frac{p_j}{2})e^{-\frac{1}{2k}y}.
\end{array}
\label{ap2}
\end{eqnarray}

\subsection*{Step 2: the solution $u(y,\tau)$ of MCH equation}

The formula (\ref{mphi}) provides a way to generate the solution $m$ of the AMCH equation from $\phi(y,\tau,\lambda)|_{\lambda=0}$.
Guided by the asymptotic behaviours (\ref{ap2}),
the $\phi$ corresponding to the boundary condition $\lim\limits_{|y|\rightarrow \infty}m(y,\tau)=k$ must be $d_1\phi_1$, where $d_1$ is a constant and $\phi_1$ is given by (\ref{mchp}).
Thus the multi-soliton solution of AMCH equation (\ref{tmCH}) reads
\begin{eqnarray}
m=\frac{\phi_1}{2\phi_{1,y}}=\frac{W_N\bar{W}_N}{2\left(W_N\bar{W}_{N,y}-W_{N,y}\bar{W}_N\right)}.
\label{mphi2}
\end{eqnarray}
Furthermore, substituting (\ref{mphi2}) into (\ref{u}), we arrive at an analytic solution of the MCH equation expressed in the variables $y$ and $\tau$
\begin{eqnarray}
u=\frac{W^3_N\bar{W}^3_N-F_N}
{2\left(W_N\bar{W}_{N,y}-W_{N,y}\bar{W}_N\right)W^2_N\bar{W}^2_N},
\label{us2}
\end{eqnarray}
where
\begin{eqnarray}
\begin{split}
F_N=&W^3_N(2\bar{W}_N\bar{W}_{N,y}\bar{W}_{N,y\tau}-2\bar{W}^2_{N,y}\bar{W}_{N,\tau}+\bar{W}_N\bar{W}_{N,\tau}\bar{W}_{N,yy}-\bar{W}^2_{N}\bar{W}_{N,yy\tau})
\\&-\bar{W}^3_N(2W_NW_{N,y}W_{N,y\tau}-2W^2_{N,y}W_{N,\tau}+W_NW_{N,\tau}W_{N,yy}-W^2_{N}W_{N,yy\tau}).
\end{split}
\label{F}
\end{eqnarray}

\subsection*{Step 3: the relation between $x$, $t$ and $y$, $\tau$}
In this step, we finish the coordinate transformation between $x$, $t$ and $y$, $\tau$.
When $\lambda=0$, the second equation in (\ref{lps}) becomes
\begin{eqnarray}
\phi_x=\frac{1}{2}\phi. \label{lps2}
\end{eqnarray}
Thus a solution may be chosen as $\phi=e^{\frac{x}{2}}$.
The solution $\phi$ of equation (\ref{lps1}) with $\lambda=0$ and $U$ chosen as in (\ref{uvmch}) is a linear combination of the two fundamental solutions $\phi_1$ and $\phi_2$ given by (\ref{mchp}).
The asymptotic behaviours (\ref{ap2}) imply that the solution corresponding to $\phi=e^{\frac{x}{2}}$ must be $d_1\phi_1$, namely,
\begin{eqnarray}
e^{\frac{x}{2}}=d_1\phi_1=d_1\frac{\bar{W}_N}{W_N}, \label{xy2}
\end{eqnarray}
where $d_1$ is an arbitrary constant. Thus we arrive at the following relation
\begin{eqnarray}
x=\ln\frac{\bar{W}^2_N}{W^2_N}+\alpha, \label{xyr2}
\end{eqnarray}
where $\alpha$ is a constant. In addition, from (\ref{rt}) it is easy to conclude that
\begin{eqnarray}
t=\tau. \label{xyr3}
\end{eqnarray}

The formulas (\ref{us2}), (\ref{xyr2}) and (\ref{xyr3}) provide an analytic solitary solution of the MCH equation in parametric form.
As examples, we present the single-soliton wave and two-soliton wave in the following.

\subsubsection*{Example 3: one-soliton solution of MCH equation}
For $N=1$, from (\ref{fj2}) we take
\begin{eqnarray}
f_1=\cosh \xi_1, \quad \xi_1=\frac{1}{2}p_1(y+c_1\tau),
\label{f1mch}
\end{eqnarray}
with $\frac{1}{4}p_1^2=-\frac{1}{4}\lambda_1^2+\frac{1}{4k^2}$, $c_1=\frac{2k^3}{k^2p_1^2-1}$.
From the general formula (\ref{uvmch}), the one-soliton solution of NKdV equation (\ref{nkdv2})  becomes
\begin{eqnarray}
\begin{split}
U&=\frac{1}{4k^2}-2(\ln f_1)_{yy}=\frac{1}{4k^2}-\frac{1}{2}p_1^2\sech^2\xi_1,
\\
V&=-2k+4(\ln f_1)_{y\tau}=-2k+c_1p_1^2\sech^2\xi_1.
\end{split}
\label{Vso}
\end{eqnarray}
Based on the general formula (\ref{mphi2}), the one-soliton solution of AMCH equation reads
\begin{eqnarray}
m=k+\frac{4k^3p_1^2}{(1-kp_1)e^{2\xi_1}+(1+kp_1)e^{-2\xi_1}+2-4k^2p_1^2}. \label{ms1}
\end{eqnarray}
From the general formulas (\ref{us2}), (\ref{xyr2}) and (\ref{xyr3}),
we arrive at the one-soliton solution of MCH equation in parametric form
\begin{eqnarray}
\begin{split}
u &=k+\frac{4k^3p_1^2\left[(1-kp_1)e^{2\xi_1}+(1+kp_1)e^{-2\xi_1}+2(1-k^2p_1^2)\right]}{(1-k^2p_1^2)\left[(1-kp_1)e^{2\xi_1}+(1+kp_1)e^{-2\xi_1}+2\right]^2},
\\
x&=\frac{y}{k}+\ln[\frac{(1-kp_1)e^{\xi_1}+(1+kp_1)e^{-\xi_1}}{2k(e^{\xi_1}+e^{-\xi_1})}]^2+\alpha,
\\
t&=\tau.
\end{split}
\label{umch1}
\end{eqnarray}

\subsubsection*{Example 4: two-soliton solution of MCH equation}

For $N=2$, we take
\begin{eqnarray}
f_1=\cosh \xi_1, \quad f_2=\sinh\xi_2, \quad \xi_1=\frac{1}{2}p_1(y+c_1\tau), \quad \xi_2=\frac{1}{2}p_2(y+c_2\tau),
\label{f2mch}
\end{eqnarray}
where $\frac{1}{4}p_1^2=-\frac{1}{4}\lambda_1^2+\frac{1}{4k^2}$, $c_1=\frac{2k^3}{k^2p_1^2-1}$, $\frac{1}{4}p_2^2=-\frac{1}{4}\lambda_2^2+\frac{1}{4k^2}$, $c_2=\frac{2k^3}{k^2p_2^2-1}$.
It follows from  the general formula (\ref{uvmch}) that the two-soliton solution of NKdV equation (\ref{nkdv2}) reads
\begin{eqnarray}
\begin{split}
U&=\frac{1}{4k^2}-2(\ln W_2)_{yy}
 =\frac{1}{4k^2}-\frac{(p_2^2-p_1^2)(p_2^2\cosh^2\xi_1+p_1^2\sinh^2\xi_2)}{2(p_2\cosh\xi_1\cosh\xi_2-p_1\sinh\xi_1\sinh\xi_2)^2},
\\
V&=-2k+4(\ln W_2)_{y\tau}
 =-2k+\frac{(p_2^2-p_1^2)(p_2^2c_2\cosh^2\xi_1+p_1^2c_1\sinh^2\xi_2)}{(p_2\cosh\xi_1\cosh\xi_2-p_1\sinh\xi_1\sinh\xi_2)^2}.
\end{split}
 \label{Usmcht}
\end{eqnarray}
Based on the general formula (\ref{mphi2}), the two-soliton solution of AMCH equation reads
\begin{eqnarray}
m=\frac{W_2\bar{W}_2}{2\left(W_2\bar{W}_{2,y}-W_{2,y}\bar{W}_2\right)},
\label{rs1mcht}
\end{eqnarray}
where
\begin{eqnarray}
\begin{split}
W_2&=\frac{1}{2}(p_2\cosh\xi_1\cosh\xi_2-p_1\sinh\xi_1\sinh\xi_2),
\\
\bar{W}_2&=\frac{1}{8}e^{\frac{y}{2k}}[p_1(p_2^2-\frac{1}{k^2})\sinh\xi_1\sinh\xi_2-p_2(p_1^2-\frac{1}{k^2})\cosh\xi_1\cosh\xi_2+\frac{1}{k}(p_1^2-p_2^2)\cosh\xi_1\sinh\xi_2].
\label{W2}
\end{split}
\end{eqnarray}
From the general formulas (\ref{us2}), (\ref{xyr2}) and (\ref{xyr3}),
we arrive at the two-soliton solution of MCH equation in parametric form
\begin{eqnarray}
\begin{split}
u&=\frac{W^3_2\bar{W}^3_2-F_2}
{2\left(W_2\bar{W}_{2,y}-W_{2,y}\bar{W}_2\right)W^2_2\bar{W}^2_2},
\\
x&=\ln\frac{\bar{W}^2_2}{W^2_2}+\alpha,
\\
t&=\tau,
\end{split}
\label{ucht}
\end{eqnarray}
where the expressions $W_2$, $\bar{W}_2$ and $F_2$ are given by (\ref{W2}) and (\ref{F}).

\section{Conclusions and discussions}

In the paper, we have proposed a simple approach to calculate multi-soliton solutions of Camassa-Holm equation and modified Camassa-Holm equation
with aid of Darboux transformation.
We first mapped the Camassa-Holm equation and modified Camassa-Holm equation to a negative order KdV equation by virtue of the reciprocal transformation.
Then by applying the Darboux transformation to the negative order KdV equation and by inverting the reciprocal transformation,
we arrived at the multi-soliton solutions of Camassa-Holm equation and modified Camassa-Holm equation.

We believe that our approach can be applied to calculate the multi-soliton solutions of other Camassa-Holm type equations,
for example, the Degasperis-Procesi equation \cite{DP}, the Novikov's cubic nonlinear equation \cite{NV1,HW1}, the two-component Camassa-Holm equations presented in \cite{CLZ}-\cite{XQZ}.
We shall study them elsewhere.

\section*{ACKNOWLEDGMENTS}
The authors Xia and Zhou were supported by the National Natural Science Foundation of China (Grant Nos. 11301229 and 11271168),
the Natural Science Foundation of the Jiangsu Province (Grant No. BK20130224) and the Natural Science Foundation of the Jiangsu Higher Education
Institutions of China (Grant No. 13KJB110009). The author Qiao was partially
supported by the National Natural Science Foundation of China (No. 11171295, No. 61301187, and No. 61328103)
and also thanks the U.S. Department of Education GAANN project (P200A120256) to support UTPA mathematics graduate program.

\vspace{1cm}
\small{

}
\end{document}